\documentclass[11pt]{article}
\usepackage{moriond,epsfig}

\usepackage{relsize}
\def\babar{\mbox{\slshape B\kern-0.1em{\smaller A}\kern-0.1em
    B\kern-0.1em{\smaller A\kern-0.2em R}}}

\def\be{\begin{equation}}
\def\ee{\end{equation}}
\def\bea{\begin{eqnarray}}
\def\eea{\end{eqnarray}}

\begin{document}
\vspace*{4cm}
\title{TEST OF LEPTON UNIVERSALITY AND SEARCHES FOR LEPTON FLAVOR VIOLATION AT BABAR}

\author{ ELISA GUIDO \\(on behalf of the \babar\ Collaboration)}

\address{Universit\`a degli Studi di Genova \& INFN Sezione di Genova\\
Via Dodecaneso 33, 16146 Genova, Italy}

\maketitle\abstracts{
The \babar\ experiment has recently obtained some important results in the search for 
new physics in leptonic and lepton flavor violating decays, exploiting the 
complete datasets collected at the $\Upsilon(4S)$, $\Upsilon(3S)$ and 
$\Upsilon(2S)$ energies. 
In particular, new limits on the ratio 
$\Gamma(\Upsilon(1S)\to\tau^+\tau^-)/\Gamma(\Upsilon(1S)\to\mu^+\mu^-)$, 
on lepton flavor violating decays of the $\Upsilon(3S)$ and $\Upsilon(2S)$, 
and on $\tau$ decays to three charged leptons or $\tau\to e/\mu \gamma$ are presented.}

\section{Introduction}

Despite being originally devoted to study CP-violation, the \babar\ experiment (described in detail elsewhere~\cite{ref:babar1,ref:babar2}) has recently obtained several important results also in precision tests of the standard model (SM) and in searches for new physics (NP) effects. This has been possible thanks to the huge data sample collected by \babar, mostly at an energy in the e$^+$e$^-$center-of-mass (CM) frame equal to the mass of the $\Upsilon(4S)$ (corresponding to an integrated luminosity of 426 fb$^{-1}$), but also of the lower-mass $\Upsilon$ resonances (28 fb$^{-1}$ at the $\Upsilon(3S)$ energy and 14 fb$^{-1}$ at the $\Upsilon(2S)$). Samples of data collected just below each $\Upsilon$ resonance (42 fb$^{-1}$, 2.4 fb$^{-1}$ and 1.3 fb $^{-1}$ below the $\Upsilon(4S)$, $\Upsilon(3S)$ and $\Upsilon(2S)$, respectively) have been used as well.

Here some recent results obtained by \babar\ are shown: a test of lepton universality in $\Upsilon(1S)$ decays~\cite{ref:LU}, representing the most precise measurement of the ratio $\Gamma(\Upsilon(1S)\to\tau^+\tau^-)/\Gamma(\Upsilon(1S)\to\mu^+\mu^-)$; and three results in the search for charged lepton flavor violating decays of the $\Upsilon(3S)$ and $\Upsilon(2S)$ resonances~\cite{ref:LFVY}, as well as of $\tau$ to $e/\mu\gamma$~\cite{ref:LFVemug} and to three charged leptons~\cite{ref:LFV3l}. 
With each result \babar\ proves to be able to constrain the NP theoretical models proposed for the different processes, as it will be explained in the devoted sections.

\section{Test of lepton universality in $\Upsilon(1S)$ decays}

In the SM, the couplings of the gauge bosons to leptons are independent of the lepton flavor. 
Aside from small lepton-mass effects, the expression for the decay
width $\Upsilon(1S)\to l^+ l^-$ should be identical for all leptons, and given by~\cite{ref:mas}:
\begin{equation}
\Gamma_{\Upsilon(1S)\to ll}=4\alpha^2 Q_b^2 \frac{|R_n(0)|^2}{M^2_\Upsilon}(1+2\frac{M^2_l}{M^2_\Upsilon})\sqrt{1-4\frac{M^2_l}{M^2_\Upsilon}}, \label{eq:BR}
\end{equation}
where $\alpha$ is the electromagnetic fine structure constant, $Q_b$ is the charge of the bottom quark, $R_n(0)$ is the non-relativistic radial wave
function of the bound $b\bar b$ state evaluated at the origin, $M_\Upsilon$ is the $\Upsilon(1S)$ mass and $M_l$ is the lepton mass. In the SM, one
expects the quantity $R_{\tau\mu}(\Upsilon(1S)) =\frac{ \Gamma_{\Upsilon(1S)\to \tau^+\tau^-}}{\Gamma_{\Upsilon(1S)\to \mu^+\mu^-}}$
to be very close to one (in particular, $R_{\tau\mu}(\Upsilon(1S))\sim0.992$~\cite{ref:PDG2008}).

In the next-to-minimal extension of the SM~\cite{ref:Higgs}, deviations of $R_{\tau\mu}$ from the SM expectation may arise due to a light CP-odd Higgs boson, $A^0$. Present data~\cite{ref:LEP} do not exclude the existence of such a boson with a mass below 10~GeV/c$^2$.
$A^0$ may mediate the following processes~\cite{ref:mas}:
\begin{equation}
\Upsilon(1S)\to A^0\gamma\to l^+l^-\gamma  \hspace{1cc} \mathrm{or}  \hspace{1cc} \Upsilon(1S)\to \eta_b(1S)\gamma, \eta_b(1S)\to A^0\to l^+l^-.\label{eqn:a0_etabmix}
\end{equation}

If the photon remained undetected, the lepton pair would be ascribed to the $\Upsilon(1S)$ and the proportionality of the coupling of the Higgs to the lepton mass would lead to an apparent violation of  lepton universality. 
The deviation of $R_{\tau\mu}$ from the expected SM value depends on $X_d=\cos\theta_A\tan\beta$ (where $\theta_A$ measures the coupling of the $\Upsilon(1S)$ to the $A^0$, and $\tan\beta$ is the ratio of the vacuum expectation values of the two Higgs doublets) and on the mass difference between $A^0$ and $\eta_b(1S)$. Assuming $X_d=12$, $\Gamma(\eta_b(1S))=5$ MeV, and the measured $M_{\eta_b(1S)}$~\cite{ref:etab}, the deviation of $R_{\tau\mu}(\Upsilon(1S))$ may be as large as $\sim4\%$, depending on the $A^0$ mass~\cite{ref:mas}.  

A measurement of this ratio has already been performed, with the result $R_{\tau\mu}(\Upsilon(1S)) = 1.02 \pm 0.02 (stat.) \pm 0.05(syst.)$~\cite{ref:CLEO}.

This analysis focuses on the measurement of $R_{\tau\mu}(\Upsilon(1S))$ in the decays $\Upsilon(3S)\to\Upsilon(1S)\pi^+\pi^-$  with $\Upsilon(1S)\to l^+l^-$ and $l=\mu,\tau$ of the $\sim1.2\times10^8$ $\Upsilon(3S)$ collected by \babar.  Only $\tau$ decays to a single charged particle (plus neutrinos) are considered, resulting in final states of exactly four detected particles for both the $\mu^+\mu^-$ and $\tau^+\tau^-$ samples.

The event selection is optimized using Monte Carlo (MC) simulated events.
Different selection criteria are used for the $\Upsilon(1S)\to\mu^+\mu^-$ decays ($D_\mu$) and  the $\Upsilon(1S)\to\tau^+\tau^-$ decays ($D_\tau$), because in the latter the presence of neutrinos in the final state leads to a larger contamination from the background (mainly non-leptonic $\Upsilon(1S)$ decays  and $e^+e^-\to\tau^+\tau^-$ events).
The final selection efficiency for the reconstructed decay chains, estimated from a sample of MC simulated events, are $\epsilon_{\mu\mu}\sim45\%$  and $\epsilon_{\tau\tau}\sim17\%$ for the $\mu^+\mu^-$ and the $\tau^+\tau^-$ final states, respectively.

\begin{figure}[!htb]
\begin{center}
\includegraphics[width=0.4\linewidth]{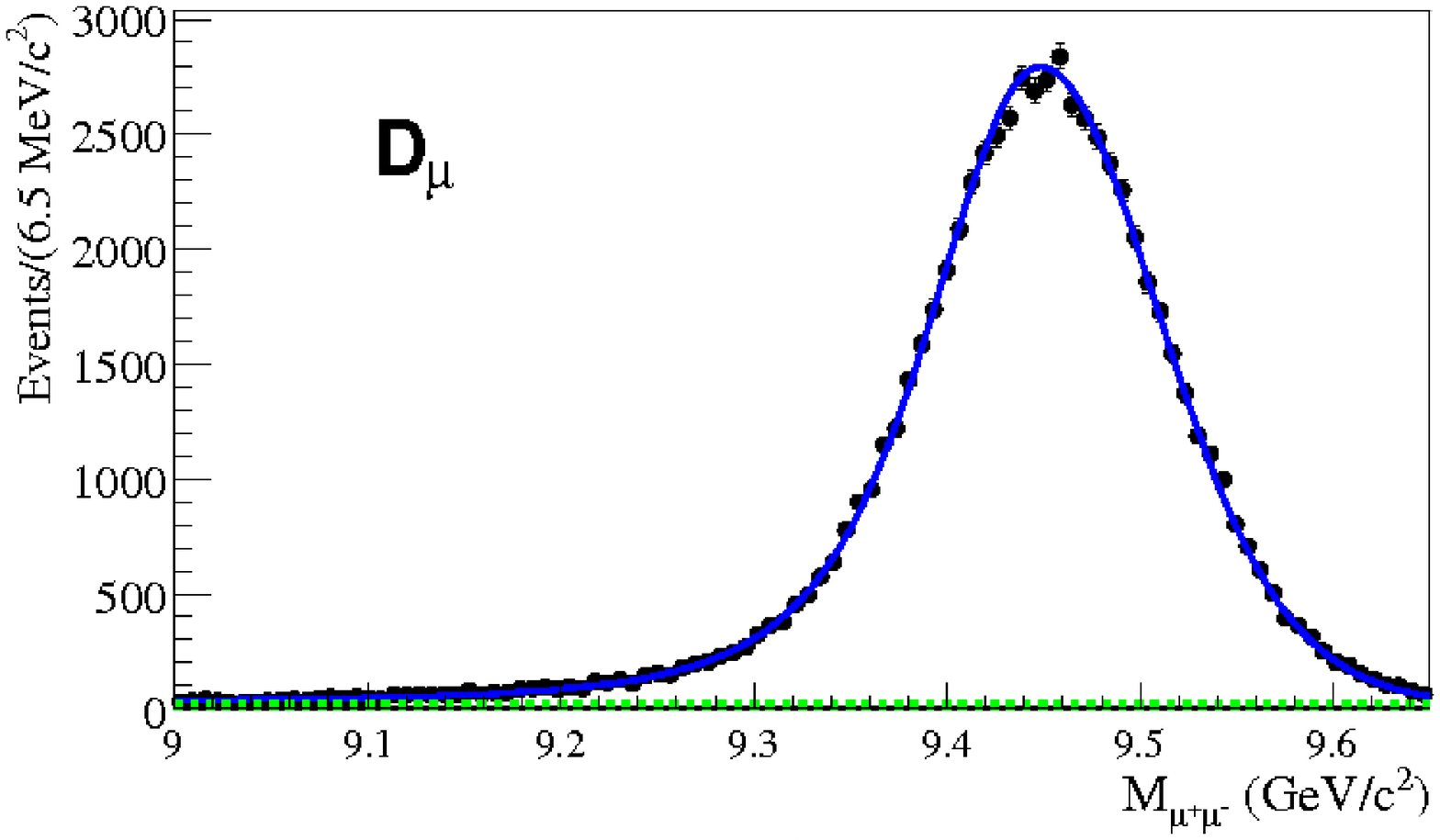}
\includegraphics[width=0.4\linewidth]{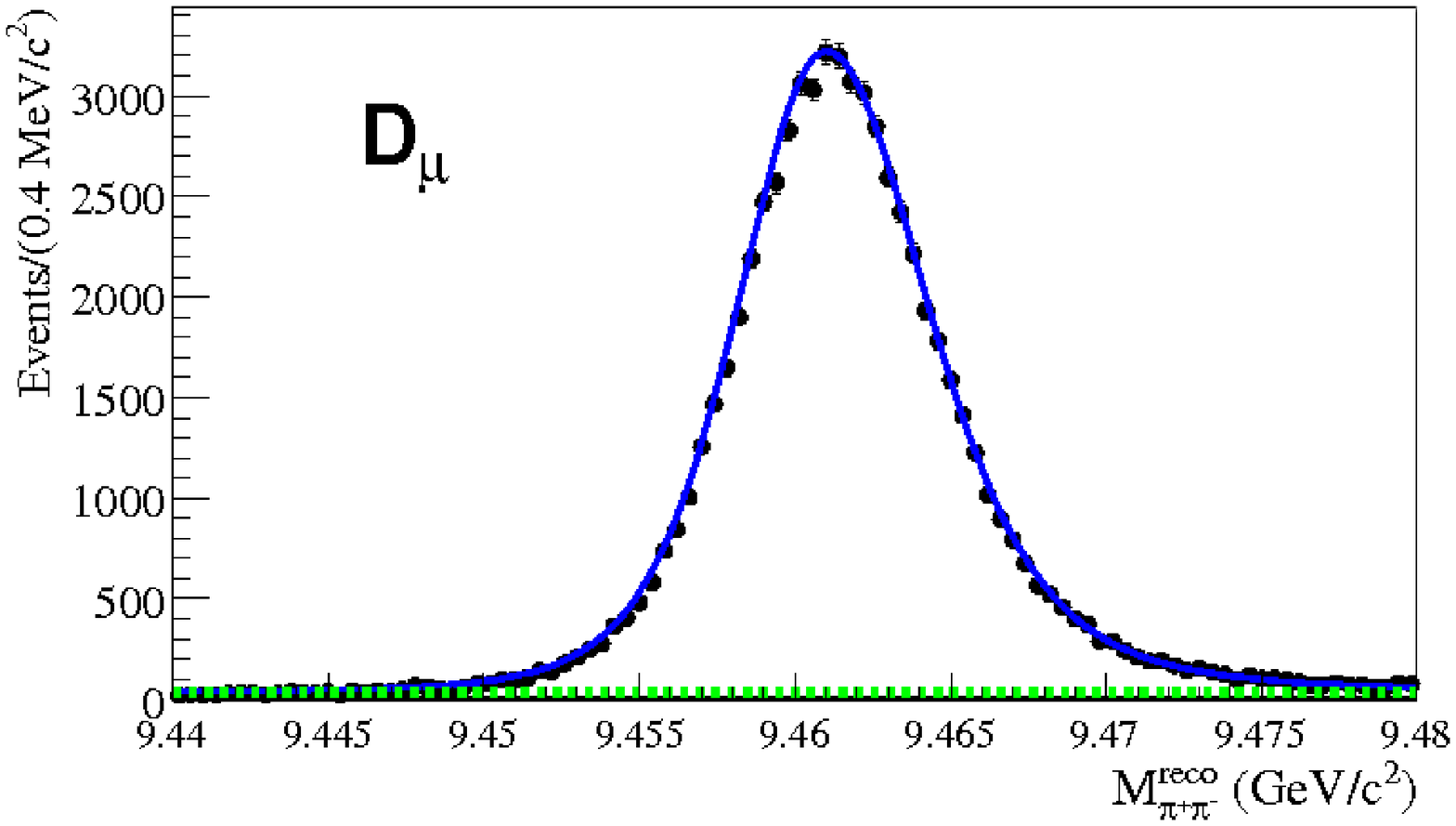}\\
\includegraphics[width=0.4\linewidth]{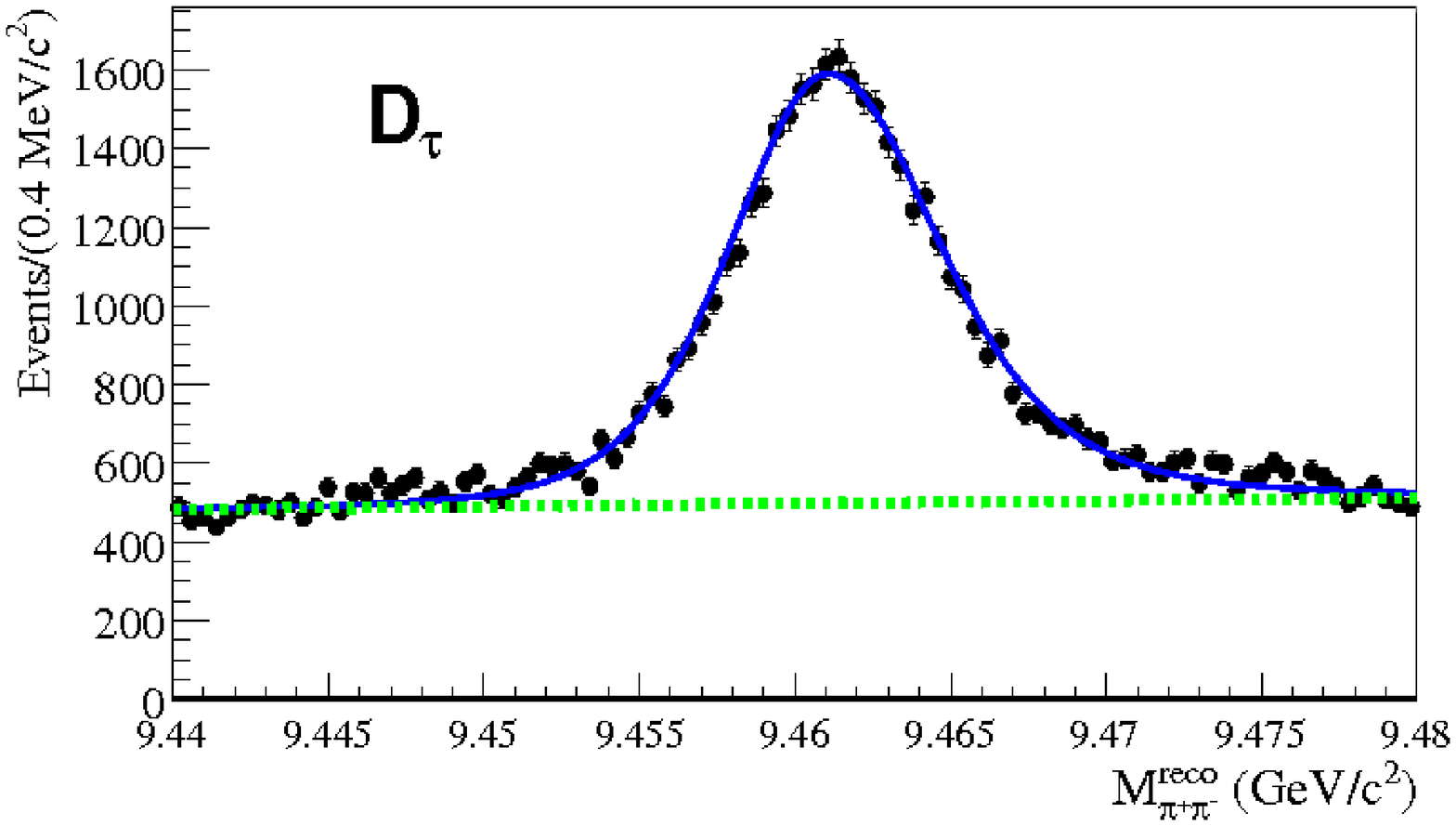}\\
\caption{1-D fit projections for $M_{\mu^+\mu^-}$ (top left) and for $M_{\pi\pi}^{reco}$ (top right) in the $D_\mu$ sample, and for $M_{\pi\pi}^{reco}$ (bottom) in the  $D_\tau$ sample. In each plot the dashed line represents the background shape, while the solid line is the sum of signal and background contributions to the fit, and the points are the data. 
}
\label{fig:LU}
\end{center}
\end{figure}

An extended unbinned maximum likelihood fit, applied simultaneously to the two disjoint datasets $D_\mu$ and $D_\tau$, is used to extract  $R_{\tau\mu}= \frac{N_{sig\tau}}{\epsilon_{\tau\tau}}\cdot\frac{\epsilon_{\mu\mu}}{N_{sig\mu}}$, where $N_{sig\mu}$ ($N_{sig\tau}$) indicates the number of signal events in the $D_\mu$ ($D_\tau$) sample. 
For the $D_\mu$ sample, a 2-dimensional probability density function (PDF) is used, based on the invariant dimuon mass $M_{\mu^+\mu^-}$ and $M_{\pi^+\pi^-}^{reco}$, the invariant mass of the system recoiling against the $\pi$-pair, defined as: $M_{\pi^+\pi^-}^{reco} = \sqrt{s+M_{\pi\pi}^2-2\cdot\sqrt{s}\cdot E_{\pi\pi}^*}$, where $\sqrt{s}$ is the $e^+e^-$ CM energy and $E_{\pi\pi}^*$ indicates the $\pi$-pair energy calculated in the CM frame.
For  the $D_\tau$ sample,  a 1-dimensional PDF is used, based on $M_{\pi^+\pi^-}^{reco}$.
The functional forms of the PDFs describing the signal components are modeled from a dedicated sub-sample consisting approximately of one tenth of the $D_\mu$ sample, then discarded from the final result in order to avoid any bias. The data collected below the $\Upsilon(3S)$ resonance are used to model the background shapes.
The result of the simultaneous fit is $R_{\tau\mu}=1.006\pm0.013$, where the quoted error is statistical only.
Figure~\ref{fig:LU} shows the projections of the fit results for the three variables.

Several systematic errors cancel in the ratio.
The main systematic uncertainties are related to the differences between data and simulation in the efficiency of event selection, 
the muon identification, and the trigger and background filters. There is also a systematic uncertainty on the signal and background yields due to the imperfect knowledge of the PDFs used in the fit.
The total systematic uncertainty, obtained by summing in quadrature all the contributions, is estimated to be $2.2\%$.
Including all the systematic corrections, the ratio $R_{\tau\mu}$ is found to be~\cite{ref:LU}:
\[
R_{\tau\mu}(\Upsilon(1S))=1.005 \pm 0.013(stat.) \pm 0.022(syst.).
\]
No significant deviation of the ratio $R_{\tau\mu}$ from the SM expectation is observed.
This result improves both the statistical and systematic precision with respect to the previous measurement~\cite{ref:CLEO}.
Assuming values for  $X_d$, $\Gamma(\eta_b(1S))$ and $M_{\eta_b(1S)}$ as previously stated~\cite{ref:mas}, the present measurement excludes an $A^0$ with mass lower than 9 GeV/c$^2$ at 90$\%$ of confidence level (CL).

\section{Searches for charged lepton flavor violation}

Lepton flavor violation (LFV) can occur via neutrino oscillation, but this has never been observed in charged processes because the tree-level contributions are suppressed to rates not achievable by the current experimental sensitivity. In many extensions of the SM, enhancements of these rates are possible, up to a detectable level, with expected branching fractions of  $\cal O$(10$^{-6}$-10$^{-8}$). An observation of LFV in charged decays would be a clear signature of NP, and improved limits on the branching fractions of such processes further constrain the theoretical models proposed.

\babar\ can search for charged LFV in several typologies of decays, both of the $\Upsilon$ resonances and of the $\tau$ leptons.

\subsection{Search for charged LFV in narrow $\Upsilon$ decays}

This analysis searches for the charged LFV decays $\Upsilon(nS)\to l^\pm\tau^\mp$, with $l=e, \mu$ and $n=2,3$, using the $\sim1.2\times10^8$ $\Upsilon(3S)$ and $\sim1.0\times10^8$ $\Upsilon(2S)$ resonances collected by \babar. 

The signature of the signal events consists of exactly two oppositely charged particles: a primary lepton, identified as an electron or a muon, with momentum close to the beam energy, and a secondary charged lepton or pion from the $\tau$ decay (along with other neutral particles not reconstructed). In order to suppress background events, if the $\tau$ decays leptonically, the primary lepton and the $\tau$-daughter are required to have different flavors. Thus, for each value of $n$, four signal channels are defined, consisting of leptonic and hadronic $\tau$ decay modes, and with an electron or a muon as primary lepton. 
The main sources of background come from $\tau$-pair production, as well as from Bhabha and $\mu$-pair events.

The event selection consists of several requirements, related to the particle identification and to the kinematics of the $\tau$-daughter. The final selection efficiencies, estimated on samples of  MC simulated events, vary in the range (4-6)$\%$ depending on the decay mode considered.

An extended unbinned maximum likelihood fit is performed to the distribution of the variable $x=|p_1|/E_B$, that is, the momentum of the primary lepton ($p_1$) normalized to the beam energy ($E_B$). The signal distribution is expected to peak at $x\sim0.97$, while the $\tau$-pair background $x$ distribution is smooth and approaches zero as $x\to0.97$, and the Bhabha and $\mu$-pair events have instead a peaking behavior at $x\sim1$. PDFs are chosen for each of these components, using samples of data and of MC simulated events. The signal yield $N_{SIG}$ is extracted and found consistent with the no signal-hypothesis within $\pm1.8\sigma$ in all the signal channels. Since no statistically significant signal is observed, the $90\%$ CL upper limit (UL) on the branching fraction $\cal B$ of each decay is determined, using a Bayesian technique, in which the prior likelihood is uniform in $\cal B$ and assumes that $\cal B$ $> 0$.

In the UL calculation, the systematic uncertainties affecting the measurement are also taken into account. The dominant contribution to the systematic uncertainty comes from an imperfect knowledge of the PDF shapes. The resulting ULs~\cite{ref:LFVY} are summarized in Table~\ref{tab:LFVY} and are of $\cal O$(10$^{-6})$, representing the first constraints on $\cal B$($\Upsilon(nS)\to e^\pm\tau^\mp$), while improving the sensitivity with respect to the previous ULs~\cite{ref:LFVYCleo} on $\cal B$($\Upsilon(nS)\to\mu^\pm\tau^\mp$).

\begin{table}[!htb]
\caption{$90\%$ CL ULs on the branching fractions $\cal B$ for signal decays $\Upsilon(nS)\to l^\pm\tau^\mp$.\label{tab:LFVY}}
\vspace{0.4cm}
\begin{center}
\begin{tabular}{|c|c|}
\hline
Mode  &  UL (10$^{-6}$)\\
\hline
$\cal B$($\Upsilon(2S)\to e^\pm\tau^\mp$)  & $<3.2$\\
$\cal B$($\Upsilon(2S)\to\mu^\pm\tau^\mp$)  & $<3.3$\\
$\cal B$($\Upsilon(3S)\to e^\pm\tau^\mp$) & $<4.2$\\
$\cal B$($\Upsilon(3S)\to\mu^\pm\tau^\mp$) & $<3.1$\\
\hline
\end{tabular}
\end{center}
\end{table}

\subsection{Search for charged LFV in the decays $\tau^\pm\to e^\pm\gamma$ and $\tau^\pm\to\mu^\pm\gamma$}

Another environment for LFV processes is $\tau$ decay. In particular, $\tau^\pm\to l^\pm\gamma$ (where $l=e,\mu$) is a favored decay mode in several NP scenarios, with predicted branching fractions close to the current experimental limits.

Besides being a $B$-factory, \babar\ has been usefully employed as a $\tau$-factory as well, since the cross sections for the production of $\tau$-pairs and $B\bar B$-pairs are comparable.
This analysis uses the complete \babar\ dataset, which corresponds to $\sim960\times10^6$ $\tau$ decays.

The reconstructed events $e^+e^-\to\tau^+\tau^-$ show a clear topology, being well divided in two hemispheres: the signal side, containing the $l^\pm\gamma$-pair, required to have mass and energy compatible with the $\tau$ mass and the beam energy, respectively; and the tag side, which is expected to contain a SM $\tau$ decay, reconstructed in events where the $\tau$ lepton goes to one or  three charged tracks (with undetected neutrinos). The signal side is further required to contain only one $\gamma$ with energy greater than 1 GeV and one track identified as an electron or a muon, separated by an angle determined by the kinematic of the process, since $\gamma$ and $l$ are emitted back-to-back in the $\tau$ rest frame.

The main sources of background come from irreducible $\tau$-pair events, $l^+l^-\gamma$ events and hadronic $\tau$ decays with mis-identification of the charged $\pi$.

Signal decays are identified by two kinematical variables: the energy difference $\Delta E=E_{l\gamma}^{CM}-\sqrt{s}/2$ and the beam-energy constrained $\tau$  mass ($m_{EC}$). The distributions of events in $m_{EC}$ versus $\Delta E$ are shown in Figure~\ref{fig:LFVemg}. In this plane, a region is defined as $m_{EC}\in[1.55,2.05]$ GeV/c$^2$ and $\Delta E\in[-0.14,0.14]$ GeV, and used to extract from fits the expected fractions of background events. The number of events in the $2\sigma$ signal ellipses ($0$ events for $\tau^\pm\to e^\pm\gamma$ and 2 events for $\tau^\pm\to\mu^\pm\gamma$) are found to be compatible with the background expectation, without evidence for a signal. After the estimate of the systematic uncertainties, which are mainly due to the efficiencies of tracking, particle identification, trigger and background filters, frequentist $90\%$ CL ULs on the branching fractions $\cal B$ of the signal processes are calculated using the POLE program~\cite{ref:POLE}. The results are~\cite{ref:LFVemug}:
\[
\cal{B}(\tau^\pm\to\mathrm{e^\pm\gamma)<3.3\times10^{-8}}\hspace{1cc} \mathrm{and} \hspace{1cc} \cal{B}(\tau^\pm\to\mu^\pm\gamma)<\mathrm{4.4\times10^{-8}}
\]
at $90\%$ CL, representing the most stringent limits on LFV in these decays.

\subsection{Limits on LFV in $\tau$ decays to three charged leptons}

A further search for LFV processes is performed by \babar\ in the neutrinoless decay $\tau^-\to l_1^-l_2^+l_3^-$, where $l_i=e,\mu$ with $i=1,2,3$, and charge-conjugate decay modes are implied throughout the section. All six lepton combinations consistent with charge conservation are considered. The data collected at the energy of the $\Upsilon(4S)$ are used, corresponding to $\sim430\times10^6$ $\tau$-pairs.

As explained for the previous analysis, $e^+e^-\to\tau^+\tau^-$ events can be clearly separated in a signal and a tag hemisphere. In this case, the signal side is required to contain three charged particles, identified as electrons or muons, according to one of the allowed combinations. In addition the three-lepton system must have mass and energy compatible with the $\tau$ mass and the beam energy, respectively. The tag $\tau$ lepton has instead to decay to one charged track, with undetected neutrinos. Therefore, the reconstructed final states consist of exactly four charged tracks with net charge equal to zero. Particle identification requirements and further selection criteria are applied in order to reject background events, which are mainly due to $q\bar q$, Bhabha and $\mu$-pair events, as well as to SM $\tau$ decays. The selection efficiencies, estimated on MC simulated samples, vary in the range (6-13)$\%$ depending on the channel considered.

\begin{figure}
 \begin{minipage}[b]{7.8cm}
  \centering
   \includegraphics[width=4.5cm]{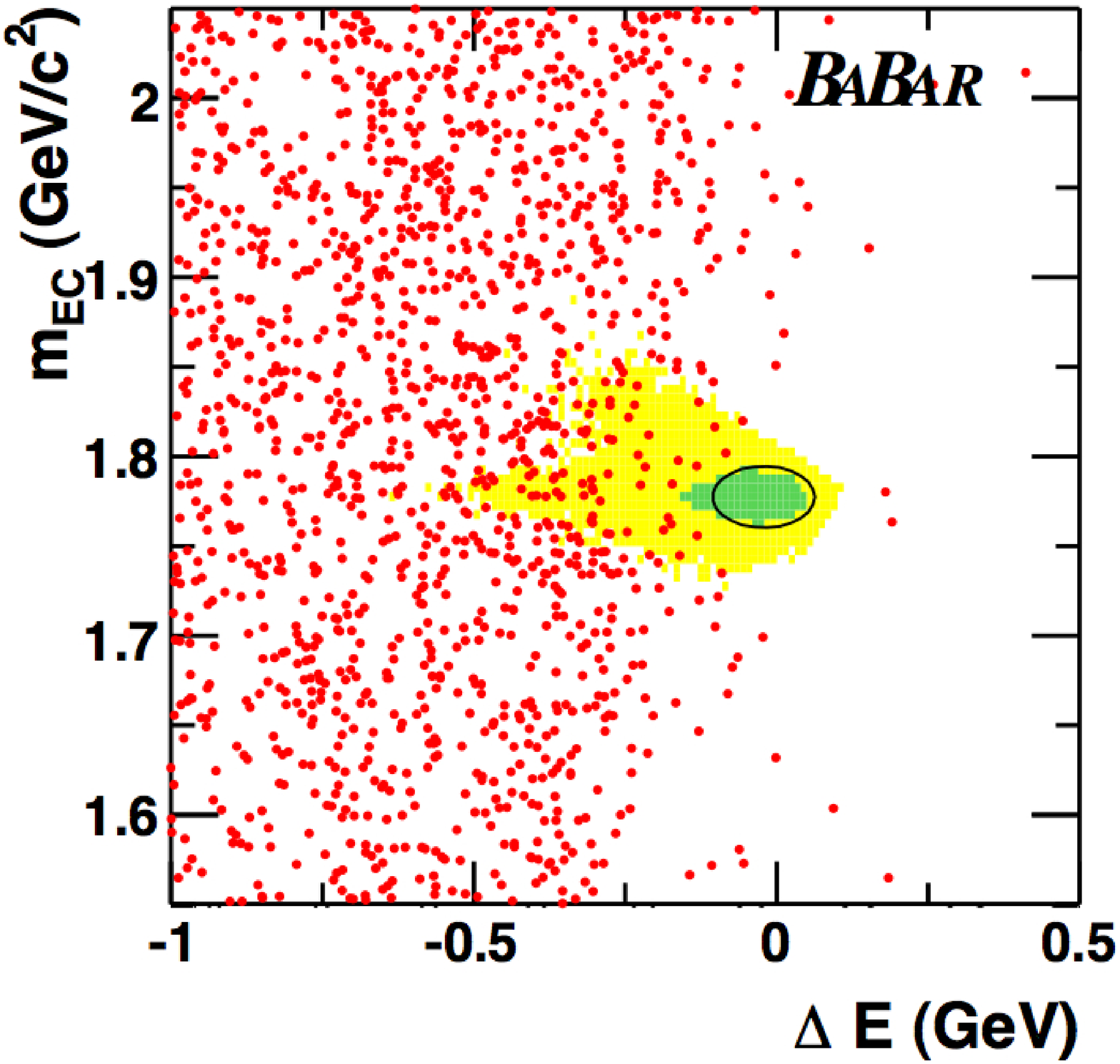}\\
   \includegraphics[width=4.5cm]{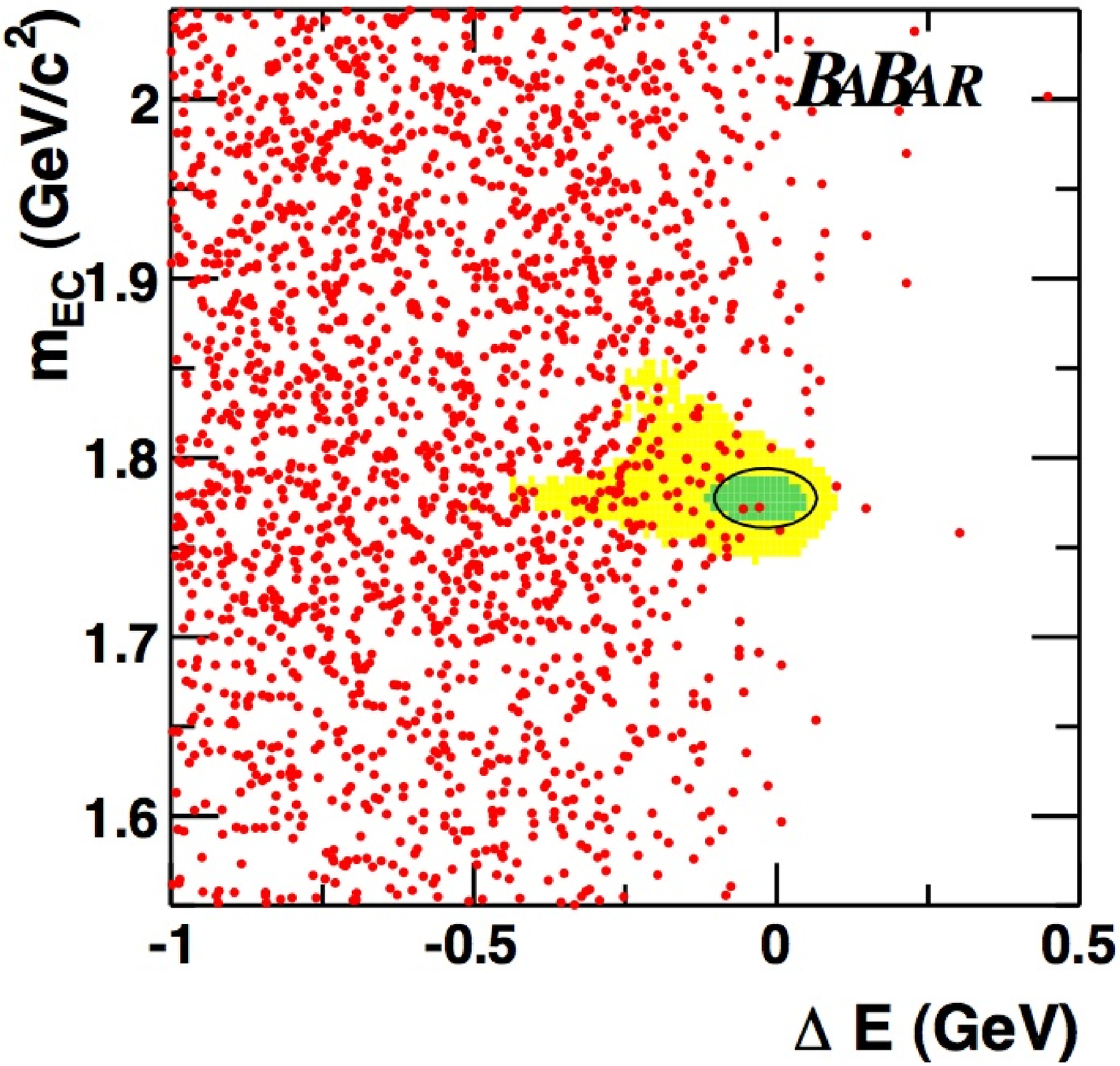}
   \caption{Data events (dots) for $\tau\to e\gamma$ (top) and $\tau\to\mu\gamma$ (bottom) decays. The $2\sigma$ signal ellipses are shown, as well as the dark and light shadings representing the $50\%$ and $90\%$ signal contours, respectively.\label{fig:LFVemg}}
 \end{minipage}
 \hspace{3mm} \
 \begin{minipage}[b]{7.8cm}
   \centering
   \includegraphics[width=6.0cm]{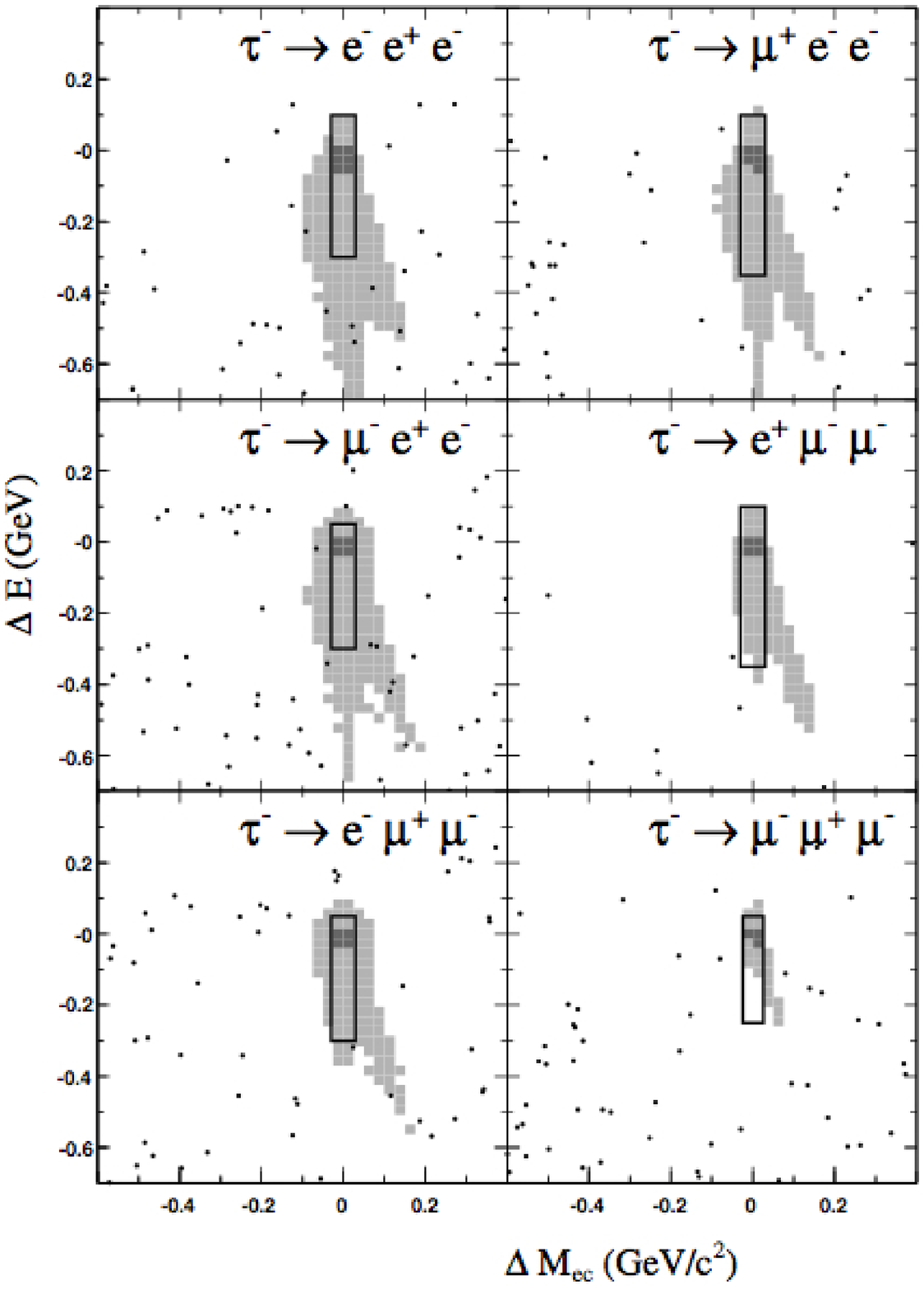}
   \caption{Data events (dots) for the six $\tau$ decay channels after selection is applied. Signal regions are identified by the solid lines. The dark and light shadings represent the 50$\%$ and 90$\%$ signal contours, respectively.\label{fig:LFV3l}}
 \end{minipage}
\end{figure}

The signal extraction is performed using the distributions $\Delta E$ and $\Delta M_{EC}=m_{EC}-m_\tau$, where $\Delta E$ and $m_{EC}$ have been defined in the previous section, and $m_\tau$ is the $\tau$ lepton mass. The distributions of the events in this plane are shown for data in Figure~\ref{fig:LFV3l}. The expected background rates for each decay mode are determined by fitting data in a region defined as $\Delta M_{EC}\in[-0.6,0.4]$ GeV/c$^2$ and $\Delta E\in[-0.7,0.4]$ GeV. In every channel no signal candidates are found in the signal region and $90\%$ CL ULs are placed on the branching fractions $\cal B$, using the technique of Cousins and Highland~\cite{ref:Cou} following the implementation of Barlow~\cite{ref:Barl}. 

The systematic uncertainties affecting the measurement are mainly due to particle identification efficiency; minor contributions come from tracking efficiency and errors in the background estimation.

After including all the uncertainties, the $90\%$ CL ULs on $\cal B$($\tau^-\to l_1^-l_2^+l_3^-$) are calculated~\cite{ref:LFV3l}, as summarized in Table~\ref{tab:LFV3l}. These values supersede the previous \babar\ results~\cite{ref:LFV3lBaBold} and are compatible with the latest limits placed by Belle~\cite{ref:LFV3lBelle}.

\begin{table}[!htb]
\caption{$90\%$ CL ULs on the branching fractions $\cal B$ for signal decays $\tau^-\to l_1^-l_2^+l_3^-$.\label{tab:LFV3l}}
\vspace{0.4cm}
\begin{center}
\begin{tabular}{|c|c|}
\hline
Mode & UL (10$^{-8}$)\\
\hline
$\cal B$($\tau^-\to e^-e^+e^-$)  & $<2.9$\\
$\cal B$($\tau^-\to\mu^-e^+e^-$) & $<2.2$\\
$\cal B$($\tau^-\to e^-\mu^+e^-$) & $<1.8$\\
$\cal B$($\tau^-\to\mu^-\mu^+e^-$) & $<3.2$\\
$\cal B$($\tau^-\to\mu^-e^+\mu^-$) & $<2.6$\\
$\cal B$($\tau^-\to\mu^-\mu^+\mu^-$) & $<3.3$\\
\hline
\end{tabular}
\end{center}
\end{table}

\newpage
\section*{References}
\begin{thebibliography}{99}

\bibitem{ref:babar1}  B.\ Aubert {\em et al.} (\babar\ Collaboration), Nucl.\ Instrum.\ Methods Phys.\ Res.\  {\bf A479}, 1 (2002).
\bibitem{ref:babar2} W.\ Menges,  IEEE Nucl.\ Sci.\ Symp.\ Conf.\ Rec.\  {\bf 5}, 1470 (2006).
\bibitem{ref:LU}  P.\ del Amo Sanchez {\em et al.} (\babar\ Collaboration), accepted by Phys.\ Rev.\ Lett., arXiv:1002.4358[hep-ex]. 
\bibitem{ref:LFVY} J.\ P.\ Lees {\em et al.} (\babar\ Collaboration), Phys.\ Rev.\ Lett.\ {\bf 104}, 151802 (2010).
\bibitem{ref:LFVemug} B.\ Aubert {\em et al.} (\babar\ Collaboration), Phys.\ Rev.\ Lett.\ {\bf 104}, 021802 (2010).
\bibitem{ref:LFV3l} J.\ P.\ Lees {\em et al.} (\babar\ Collaboration), submitted to  Phys.\ Rev.\ D(RC), arXiv:1002.4550[hep-ex].

\bibitem{ref:mas} M.~A.~Sanchis-Lozano, Int. J. Mod. Phys. {\bf A19}, 2183  (2004);  E.~Fullana and M.~A.~Sanchis-Lozano, Phys.\ Lett.\  {\bf  B653}, 67 (2007); F.~Domingo {\it et al.}, JHEP01 (2009) 061.
\bibitem{ref:PDG2008}  C.~Amsler {\it et al.}  (Particle Data Group),  Phys.\ Lett.\  {\bf B667}, 1 (2008).
\bibitem{ref:Higgs} R.~Dermisek and J.~F.~Gunion, Phys.\ Rev.\ Lett.\  {\bf 95}, 041801 (2005).  
\bibitem{ref:LEP} S.~Kraml {\it et al.}, CERN 2006-009; S.~Schael  {\it et al.} (ALEPH, DELPHI, L3 and OPAL Collaborations), Eur. Phys. J. {\bf C47}, 547 (2006).
\bibitem{ref:etab}  B.~Aubert {\it et al.}  (\babar\ Collaboration), Phys.\ Rev.\ Lett.\ {\bf 101}, 071801 (2008).
\bibitem{ref:CLEO}  D.~Besson {\it et al.} (CLEO Collaboration), Phys.\ Rev.\ Lett.\  {\bf 98}, 052002 (2007).


 
\bibitem{ref:LFVYCleo} W.~Love {\it et al.} (CLEO Collaboration), Phys.\ Rev.\ Lett.\ {\bf 101}, 201601 (2008).

\bibitem{ref:POLE} J.~Conrad {\it et al.}, Phys.\ Rev.\ {\bf D67}, 012002 (2003).

\bibitem{ref:Cou} R.~D.~Cousins and V.~L.~Highland, Nucl.\ Instrum.\ Methods Phys.\ Res.\ {\bf A320}, 331 (1992).
\bibitem{ref:Barl} R.~Barlow, Comput.\ Phys.\ Commun.\ {\bf 149}, 97 (2002).
\bibitem{ref:LFV3lBaBold} B.~Aubert {\it et al.} (\babar\ Collaboration), Phys.\ Rev.\ Lett.\ {\bf 99}, 251803 (2007).
\bibitem{ref:LFV3lBelle} K.\ Hayasaka {\it et al.} (Belle Collaboration), submitted to Phys.\ Lett.\ B, arXiv:1001.3221[hep-ex].




\end{thebibliography}
\end{document}